\documentclass[prb,twocolumn,showpacs,preprintnumbers,amsmath,amssymb]{revtex4}
\usepackage{graphicx}
\usepackage{color}

\newcommand{\muB}{\ensuremath{\mu_{\mathrm{B}}}}
\newcommand{\Tc}{\ensuremath{T_{\mathrm{C}}}}
\newcommand{\EF}{\ensuremath{E_{\mathrm{F}}}}

\begin{document}

\title{Ferromagnetism in Nitrogen-doped MgO}

\author{Phivos Mavropoulos}\email{Ph.Mavropoulos@fz-juelich.de} 
\author{Marjana Le\v{z}ai\'{c}}\email{M.Lezaic@fz-juelich.de}
\author{Stefan Bl\"ugel}

\affiliation{Institut f\"ur Festk\"orperforschung (IFF) and Institute for Advanced Simulation (IAS), Forschungszentrum
  J\"ulich, D-52425 J\"ulich, Germany}

\begin{abstract}
  The magnetic state of Nitrogen-doped MgO, with N substituting O at
  concentrations between 1\% and the concentrated limit, is calculated
  with density-functional methods. The N atoms are found to be
  magnetic with a moment of 1~$\mu_B$ per Nitrogen atom and to
  interact ferromagnetically via the double exchange mechanism. The
  long-range magnetic order is established above a finite
  concentration of about 1.5\% when the percolation threshold is
  reached. The Curie temperature $T_\mathrm{C}$ increases linearly
  with the concentration, and is found to be about 30~K for 10\%
  concentration. Besides the substitution of single Nitrogen atoms,
  also interstitial Nitrogen atoms, clusters of Nitrogen atoms and
  their structural relaxation on the magnetism are discussed. Possible
  scenarios of engineering a higher Curie temperature are analyzed,
  with the conclusion that an increase of $T_\mathrm{C}$ is diﬃcult to
  achieve, requiring a particular attention to the choice of
  chemistry.
\end{abstract}

% 75.50.Pp: Magnetic semiconductors
% 75.30.Hx: Magnetic impurity interactions
% 75.30.Et: Exchange and superexchange interactions

\maketitle

\section{Introduction}

In the research field of diluted magnetic semiconductors (DMS), a new
direction is being investigated in the last five years, namely the
engineering of ferromagnetic state formation by $sp$ impurity
doping. Compared to the more traditional DMS with transition-metal
impurities, the novel $sp$-magnetism, or $d^0$-magnetism, is rather
unexplored. The increasing interest is due to the perhaps unexpected
finding that $p$-bands can spontaneously polarize giving a
ferromagnetic state (although the possibility of magnetic $sp$-defects
is long known), but also due to the hope of tuning the properties of
these states so that high Curie temperatures are achieved even at low
concentrations.

\begin{figure}
\begin{center}
\includegraphics[width=8cm]{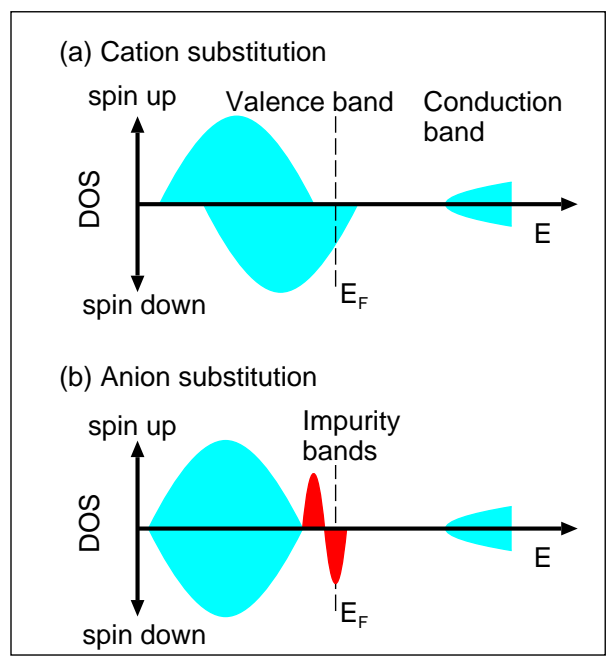}
\end{center}
\caption{(color online) Typical mechanisms for $sp$ ferromagnetism:
  (a) cation substitution drives the Fermi level \EF\ into the
  valence band; (b) anion substitution creates an impurity band in
  the gap.\label{fig:1}}
\end{figure}

Theoretical considerations have revealed mainly two routes for the
formation of $sp$-ferromagnetic states. In the first, the
semiconductor or insulator cation is substituted by an atom of smaller
valency, thus depriving the $p$-type valence band from electrons. This
hole-doping can shift the Fermi level into the valence band deep
enough that the Stoner criterion is fulfilled, and spontaneous spin
polarization appears (see Fig.~\ref{fig:1}a). This is, for example,
the mechanism encountered in alkali-atom doped TiO$_2$ and ZrO$_2$,
predicted to be ferromagnetic by {\em ab-initio}
calculations.\cite{Maca08} The hole-doping can also be achieved by
cation vacancies, instead of substitution, as was e.g. proposed for
the cases of HfO$_2$, CaO, and ZrO$_2$.\cite{Pemmaraju05,Osorio05,Bouzerar06}

The second scenario is that the anion is substituted by one of smaller
valency, introducing shallow, spin-polarized gap states. As the
impurity concentration increases, these states form impurity bands,
which remain spin-polarized if the Stoner criterion is fulfilled (see
Fig.~\ref{fig:1}b). The magnetic moment is expected to be strongest
for $2p$ impurities, i.e., in the case in Carbon- or Nitrogen-doped
oxides.\cite{Kenmochi04,Kenmochi04b,Dinh05,Elfimov07,Bannikov07,Droghetti08,Pardo08,Gu09,Droghetti09}
This is because the $2p$ states have no nodes and are rather
localized, leading to a significant Hund's-type exchange interaction
on-site.  In the concentrated limit (full substitution of the anion,
see e.g.\ Ref.~\onlinecite{Sieberer06}), both routes converge to the
same mechanism.

While the appearance of $p$-type magnetic states in oxides has been
studied in the past, preliminary predictions of the Curie temperature
have been so-far based only on mean-field
theory,\cite{Kenmochi04,Kenmochi04b,Pardo08} which has since been
shown to give qualitatively wrong results, overestimating \Tc, because
it ignores the fundamental phenomenon of magnetic percolation in
diluted disordered magnetic
systems.\cite{Sato04,Bergqvist04,Hilbert04} This means that, at low
concentrations, the average inter-impurity distance is large, so that
the short-range exchange interactions cannot produce a high Curie
temperature although they are strong.

The scope of this paper is to investigate the appearance and stability
of the ferromagnetic state, including calculations for the Curie
temperature, in MgO$_{1-x}$N$_x$ compounds for $x<15\%$, in view of
recent experimental activities in the particular system.\cite{Parkin}
Our focus is on the solution provided by local density-functional
theory (DFT), presented in Sec.~\ref{sec:groundstate}, and on the
calculation of the Curie temperature (Sec.~\ref{sec:tc}) with the
exchange constants harvested within the adiabatic approximation. As it
turns out, for any reasonable concentration the Curie temperatures are
significantly lower than room temperature, therefore, we also explore
the possibility of increasing the Curie temperature by heavy doping in
Sec.~\ref{sec:change}. Further on, we discuss the physics that lies
beyond our approximations in Sec.~\ref{sec:approximations}. There,
among our considerations, a comparison is made to previously
calculated results for N-doped MgO that focus on the effects of
electron correlation, and we also explore the constraints imposed by
our specific structural model, including a discussion on the
solubility of N in MgO. Finally, we give an outlook in
Sec.\ref{sec:conclusions}. The methods of calculation are shortly
described in the Appendix.

\section{Ground-state electronic and magnetic structure\label{sec:groundstate}}

MgO crystallizes in the rock-salt structure with a lattice parameter
of 4.21 \AA\ and exhibits a wide band gap of 7.8~eV. In our
density-functional calculations the gap is found to be 4.8~eV, due to
the well-known underestimation of insulator band gaps within local
density-functional theory [local density approximation (LDA) or
generalized gradient approximation (GGA)]. When substituting O, N
induces $p$ states in the MgO gap, approximately 0.5~eV above the
valence band edge. As N has one electron less than O, the gap states
host one unpaired hole, showing a magnetic moment of 1~\muB\ per N
atom.

At finite concentrations, interaction among the impurity states forms
a partially filled impurity band. The spin polarization remains
present even at high concentrations, showing that magnetism is not
only a consequence of the missing electron, but also of the relatively
strong Hund's exchange due to the localization of the $2p$ states.
The spin-polarized density of states shows a half-metallic behavior,
as we found by calculations with the Korringa-Kohn-Rostoker Green
function method\cite{SPRTBKKR} (KKR) within the Coherent Potential
Approximation to disorder (CPA). (Details on the calculation methods
are given in the Appendix.) Through a calculation of structural
relaxation around a N impurity, performed with the VASP\cite{VASP}
projector augmented wave code at 5\% concentration in supercell
geometry, we could rule out serious Jahn-Teller distortions around the
impurity: the first neighbors relax outward by less that 2\%,
including only a weak symmetry breaking along one of the three
crystallographic axes. This leads to a small lifting of degeneracy of
the impurity state, insignificant compared to the impurity band
width. The small atomic displacements also justify the use of CPA.

\begin{figure}
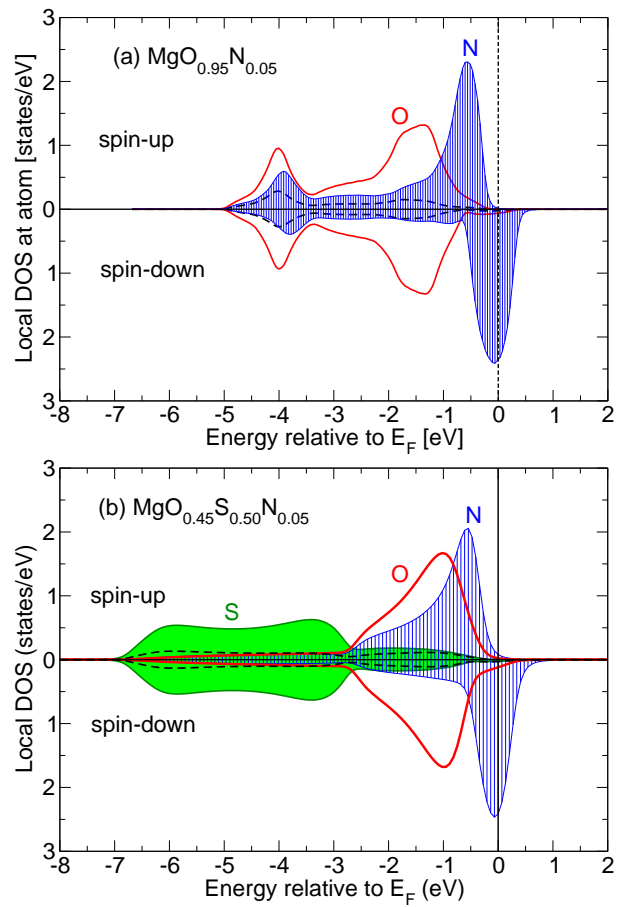

\begin{center}
\includegraphics[width=8cm]{dos_MgON.eps}
\includegraphics[width=8cm]{dos_MgOSN.eps}
\caption{(color online) (a) Atom-resolved local densities of states
  (LDOS) of MgO$_{0.95}$N$_{0.05}$ calculated within the KKR-CPA. The
  LDOS of the Mg atom is presented by a dashed line. (b) Same for
  MgO$_{0.45}$S$_{0.5}$N$_{0.05}$.\label{fig:dos}}
\end{center}
\end{figure}

We turn now to the discussion of the electronic and magnetic
structure.  Fig.~\ref{fig:dos}(a) shows the atom-resolved local
densities of states (DOS) at a N concentration of $x$=5\%. Evidently N
has a spin-split DOS and exhibits a local magnetic moment. The N
majority-spin (spin-up) impurity band is fully occupied and the
minority-spin (spin-down) impurity band is occupied by 2/3. As a
consequence, 1/3 the total moment is 1~\muB\ per N atom. At lower
concentrations the DOS is very similar, but with smaller impurity band
width $w$ ($w$ depends on the concentration as $w\sim\sqrt{x}$, since
$x$ represents an average number of impurity neighbors, and it is
known from the tight-binding approximation that $w$ increases as the
square root of the number of neighbors). The exchange splitting is of
the order of 0.5~eV, giving an exchange integral of $I=0.5~{\rm
  eV}/\muB$, which is approximately half the one in $3d$ transition
metals. 

Already from examining the ferromagnetic density of states
it is expected that the ferromagnetic state will be stable, since \EF\
lies within the impurity band, favoring
double-exchange\cite{Akai98,Sandratskii06} (i.e., upon ferromagnetic
alignment, hybridization leads to a broadening of the partially-filled
impurity-band, resulting in energy gain). We confirmed this by
calculations (not shown here in detail) of the total energy of the
ferromagnetic state, $E_{\rm ferro}$, versus the
disordered-local-moment state energy, $E_{\rm DLM}$; the latter is
represented in the CPA by an alloy of the type
MgO$_{1-x}$N$^{\uparrow}_{0.5x}$N$^{\downarrow}_{0.5x}$, where
N$^{\uparrow}$ and N$^{\downarrow}$ are impurities with magnetic
moment pointing ``up'' and ``down''.\cite{Akai98} As is typical for
double-exchange DMS systems,\cite{Sato04} we found that the energy
gain of the ferromagnetic state, $E_{\rm DLM}-E_{\rm ferro}$, scales
with the square root of the concentration. Further confirmation on the
ferromagnetic nature of the ground state comes from calculations of
the exchange constants, to be presented in the next Section.

\begin{figure}
\begin{center}
\includegraphics[width=8cm]{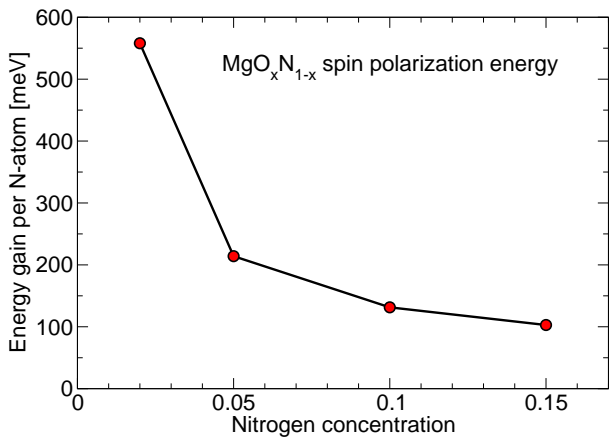}
\caption{Spin-polarization energy per Nitrogen atom as a function of
concentration $x$. As the concentration increases, the hybridization
increases, the impurity band becomes more itinerant, and the
spin-polarization energy per atom drops. The line is a guide to the eye.\label{fig:energy}}
\end{center}
\end{figure}

\begin{figure}
\begin{center}
\includegraphics[width=8cm]{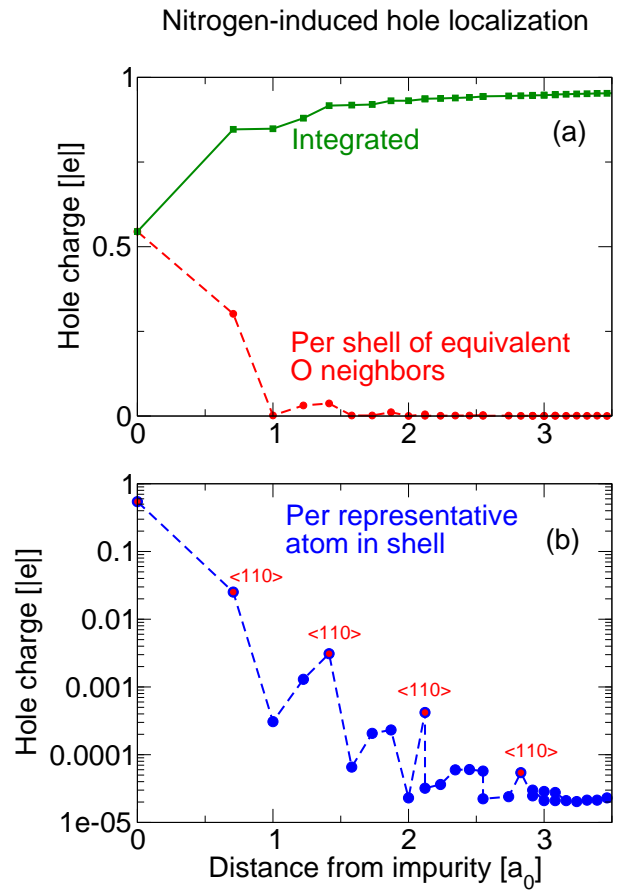}
\caption{(color online) Spatial distribution of the Nitrogen-induced
  spin-polarized hole for the case of a single Nitrogen impurity
  substituting O in MgO. Only the fraction of the hole at the N and O
  atoms is presented (the fraction at the Mg atoms is negligible). The
  distribution of the magnetic moment (1 \muB) is practically the
  same. (a) Full (green) curve: Integrated hole charge as a function
  of distance from the impurity. Dashed (red) curve: Hole charge per
  shell of O atoms that are equidistant from N and equivalent by the
  point-group symmetry. The impurity accommodates more than half of the
  hole charge. (b) Hole charge per atom as a function of distance from
  the impurity. Only one of the symmetry-equivalent atoms is
  considered for every shell. The points labelled $\langle 110
  \rangle$ represent atoms in the $\langle 110 \rangle$ direction, in
  which the extent of the impurity wavefunction is largest.  ${\rm
    a}_0$ is the lattice parameter.\label{fig:hole}}
\end{center}
\end{figure}

The spin polarization energy (i.e., the energy gain of the system due
to the moments' formation) drops with concentration due to the
increase of hybridization, ranging between 550 and 100~meV for
$2\%<x<15\%$, as can be seen in Fig.~\ref{fig:energy}. However, a high
spin-polarization energy is a necessary but not sufficient condition
for the stability of the ferromagnetic state and for a high Curie
temperature. The extension of the impurity gap-state is also
important, as it dictates the ``communication'' between localized
moments and determines the onset of magnetic
percolation.\cite{Sato04,Bergqvist04} We therefore show in
Fig.~\ref{fig:hole}(a) the extension of the Nitrogen-induced
spin-polarized hole (calculations here were done within the KKR
impurity-in-host approach, assuming that the three N-induced gap
states are degenerate and occupied by 2/3 each).  We find that
approximately half of the hole is localized at the impurity site,
while most of the remaining charge is distributed at the 12 nearest
Oxygen neighbors. From Fig.~\ref{fig:hole}(b) it can also be seen that
the hole falls off exponentially with distance, showing, however, a
higher value in the $\langle 110\rangle$ directions. These are the
directions of the first Oxygen neighbors in the fcc lattice (which is
the sublatice formed by Oxygen atoms in MgO), and in these directions
also the pair exchange constants fall off slower (see
Sec.~\ref{sec:tc}). The hole occupation at Mg sites is negligible.

\section{Exchange interactions and Curie temperature \label{sec:tc}}

\begin{figure}
\begin{center}
\includegraphics[width=8cm]{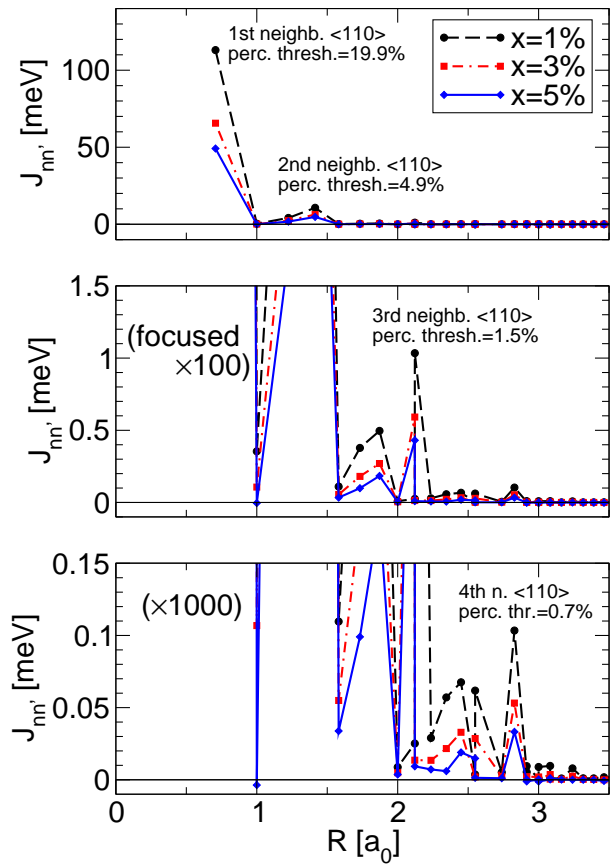}
\caption{(color online) Pair exchange interactions $J_{nn'}$ as a
  function of distance $R$ between nitrogen atoms for $x=1\%$, 3\%,
  and 5\%. The peaks corresponding to neighbors in the $\langle
  110\rangle$ directions are indicated, together with the percolation
  threshold for interactions up to the particular distance. ${\rm
    a}_0$ is the lattice parameter. The lines are guides to the
  eye. \label{fig:Jij}}
\end{center}
\end{figure}

We turn now to the discussion of the Curie temperature. We describe
the fluctuations of the magnetization on the basis of the classical
Heisenberg Hamiltonian,
$H=-\sum_{nn'}J_{nn'}\,\hat{e}_n\cdot\hat{e}_{n'}$. Here, $J_{nn'}$
are the pair exchange parameters between N impurities at sites $n$,
$n'$ and $\hat{e}_{n}$, $\hat{e}_{n'}$ are unit vectors pointing in
the direction of the local moments. The Heisenberg model includes the
transversal degrees of freedom of the fluctuating magnetic moments. As
we see below, this energy scale is much smaller than the
spin-polarization energy, and thus the longitudinal fluctuations of
the magnetic moments are safely ignored as regards the Curie point.

The $J_{nn'}$ are fitted to LDA results so that at the end the
magnetic excitation energies of the Heisenberg Hamiltonian correspond
to the ones of MgO$_{1-x}$N$_x$. For the fitting we employ the method
of infinitesimal rotations.\cite{Liechtenstein87} The results are
shown in Fig.~\ref{fig:Jij}. Clearly, all interactions are
ferromagnetic. We note that there is a marked concentration
dependence, with stronger exchange constants at low concentrations, as
is well-known in DMS systems with double-exchange ferromagnetic
interactions (see, for example, Ref.~\onlinecite{Belhadji07}).

Note, however, that there is a ``hidden'' antiferromagnetic
superexchange interaction, reducing the value of $J_{nn'}$. This
arises from the proximity of the impurity bands of both spins to
\EF: in an antiferromagnetic alignment of two impurities, energy is
gained due to hybridization of majority-spin states of each impurity
with minority-spin states of the other, accompanied by downward
shifting of the occupied majority-spin bands and upward shifting of
the minority-spin bands.\cite{Belhadji07} This superexchange
counter-acts the double exchange to some extent. To reveal this, we
performed a non-self-consistent calculation of $J_{nn'}$ (just one
iteration) after shifting the majority-spin potential of N to lower
energies by 2.7~eV. In this way the majority-spin N band is driven
away from the Fermi energy and superexchange is strongly reduced. The
result (not shown here) was striking: the $J_{nn'}$ became stronger by
approximately 50\% even at long distances.

It is evident that the nearest-neighbor exchange constants are large
(Fig.~\ref{fig:Jij}), though insignificant, in the dilute case, for
the stability of the ferromagnetic state; this is governed by
longer-range exchange, due to the requirement for magnetic
percolation.\cite{Sato04,Bergqvist04} However, the $J_{nn'}$ fall off
exponentially with distance, as is expected by the fact that the Fermi
level falls in the gap at least for the one spin
direction\cite{Pajda02} (half-metallic or insulating
systems). Notably, the slowest decay of $J_{nn'}$ is observed along
$\langle 110\rangle$, i.e., in the directions where the charge of the
impurity gap-state falls off with the slowest rate (see also
Fig.~\ref{fig:hole}). Note that the interactions in the $\langle
110\rangle$ directions are also dominating in zinc-blende or
diamond-structure DMS with transition-element impurities.\cite{Sandratskii03,Picozzi06}

\begin{figure}
\begin{center}
\includegraphics[width=8cm]{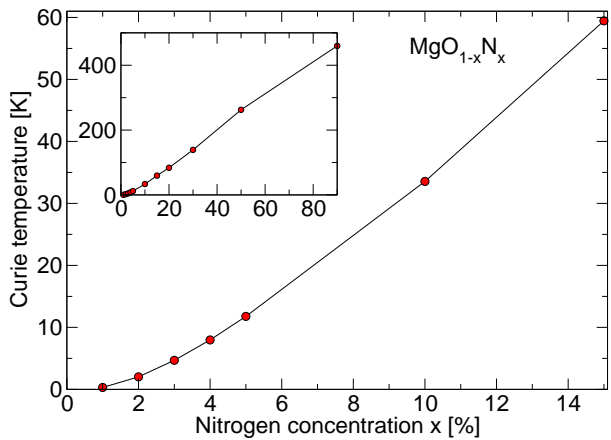}
\caption{Curie temperatures for N concentrations between 1\% and 15\%,
  calculated within the RPA. Inset: Same in the concentration range
  below 1-90\%. The lines are guides to the eye.\label{fig:Tc}}
\end{center}
\end{figure}

In Fig.~\ref{fig:Jij} we also give the percolation thresholds which we
calculated for interaction distances corresponding to ``peaks'' of
$J_{nn'}$ in the $\langle 110\rangle$ directions. We see that the
nearest-neighbor coupling starts playing a role at $x=20\%$, while the
next peak, corresponding to $R = \sqrt{2}\,{\rm a}_0$, becomes
important only for concentrations above $x=4.9\%$; at this
concentration, $J(\sqrt{2}\,{\rm a}_0)\approx 5~{\rm meV}$. At
concentrations of the order of $1.5\%$, where we see the onset of the
Curie temperature in Fig.~\ref{fig:Tc} (discussed in more detail
below), the third neighbor in the $\langle 110\rangle$ direction
becomes important, with a low value of $J(\frac{\sqrt{3}}{2}\,{\rm
  a}_0)\approx 1~{\rm meV}$. We consider interactions beyond this
distance as negligibly small, therefore we recognize the concentration
of 1.5\% as a magnetic percolation threshold for MgO$_{1-x}$N$_x$;
this can be seen also from the concentration-dependent Curie
temperature, as we discuss below. (Of course $J(R)$ is never {\em
  exactly} zero, therefore the choice of $x=1.5\%$ as percolation
threshold is somewhat arbitrary, but, we think, reasonable.)

A strong exponential decrease of $J_{nn'}$ is an indication of a low
Curie temperature in diluted magnetic systems. We have calculated
$T_{\rm C}$ within the random-phase approximation for disordered
systems, following the prescription of Hilbert and
Nolting\cite{Hilbert04} (see the Appendix for calculational
details). In doing this, we assumed the limit of classical spins
($S\rightarrow\infty$, with appropriate re-normalization of the
exchange constants), for reasons that we discuss in
Sec.~\ref{sec:approximations}. In special cases we also performed
Monte Carlo calculations, as described in the Appendix, which gave
typically 50\% higher \Tc. The results on the Curie temperature, as a
function of concentration, are shown in Fig.~\ref{fig:Tc}. Although in
this work we are interested in concentrations up to the order of
$x=10$-15\%, in the calculation we included also high concentrations,
in order to see the trend when approaching the concentrated
limit. Starting from a percolation threshold at approximately
$x=1.5$\%, $T_{\rm C}$ rises linearly with concentration. Although, as
we saw, the exchange constants at any particular distance weaken with
increasing concentration, the important effect here comes from the
decrease of the average inter-impurity distance at higher $x$,
allowing for the shorter-ranged, stronger interactions to play a role
in the ferromagnetic ordering.

At concentrations that are usually considered in DMS systems, i.e.,
not higher than 10-15\%, the Curie temperature turns out to be rather
low: we find $T_{\rm C}=35$~K at 10\%, and even at $x=20$\%, $T_{\rm
  C}$ is still below 100~K. This is a clear drawback for application
purposes at room temperature, and in the following Section we consider
possible ways to overcome this difficulty.

\section{Considerations on how to increase $T_{\rm C}$ \label{sec:change}}

Since the Curie temperature of diluted magnetic systems depends on the
long-range exchange constants, we have examined several possibilities
of increasing $J_{nn'}$ at distances larger than nearest
neighbors. Considering that $J_{nn'}$ falls off exponentially with
distance $R_{nn'}=|\vec{R}_n-\vec{R}_{n'}|$ with a characteristic
decay parameter $\kappa$, $J_{nn'}\sim \exp(-\kappa R_{nn'})$, the
main scope is to reduce $\kappa$. We tried to do this in several ways,
as we discuss here, however, none produced a significant improvement.

--\emph{Reducing the band gap.} Perhaps the most obvious way to reduce
$\kappa$ is to engineer a smaller gap by alloying. For instance,
Mg$_{1-x}$Zn$_x$O shows a smaller gap, while retaining the rock-salt
structure at not too high Zn concentrations; numerous other alloying
combinations could have a similar effect. We pursued this idea by
acting with an attractive constant potential on the Mg site, thus
artificially reducing the MgO gap size. We then derived $\kappa(E)$ by
calculating the complex band structure for energies in the gap region,
i.e., looking for solutions of the Schr\"odinger equation that fall
off as $\psi\sim e^{-\kappa r}$ (this procedure is well-known in the
calculation of surface and interface states). For $E$ deep in the gap
there is a significant reduction of $\kappa(E)$ with the gap
size. However, for $E$ at the actual position of the N levels
(relatively close to the valence band edge $E_v$), $\kappa(E)$ is
mainly determined by the effective mass $m^*$ of the light-hole
valence band as $\kappa\approx\sqrt{2m^*(E-E_v)}/\hbar$, due to the
analytical behavior of the complex band structure. (Note that, among
the complex bands, we seek the one with smallest value of $\kappa$,
which corresponds to the band derived from the light
holes.\cite{Mavropoulos00}) As the effective mass does not change
appreciably, we find that a reduction of the gap does not affect
$\kappa(\EF)$ significantly, thus it cannot change the long-range
exchange coupling.

--\emph{Changing the effective mass by compression.} After the
previous considerations, the next obvious idea is to reduce the
effective mass by compression, to be achieved, e.g., by epitaxial
stress or strain. Calculations for a 5\% reduction of the lattice
constant show that only the effective mass of the heavy holes is
appreciably affected, while the light hole bands are not much
altered. However, what is important for a small $\kappa$ is the
light-hole behavior; thus $\kappa(\EF)$ does not change practically,
and there is no increase in $J_{nn'}$. Actually for large distances we
do obtain small changes in $J_{nn'}$ after compression, but towards
antiferromagnetic behavior. This has probably two
origins:\cite{Bergqvist08} at smaller lattice constants, the band
width increases and the exchange splitting becomes smaller, so that
(i) the spin-down impurity band merges more with the valence band
setting on a Ruderman-Kittel-Kasuya-Yosida (RKKY) behavior, and (ii)
the antiferromagnetic superexchange, which is sensitive to the
exchange splitting, increases.

--\emph{Alloying with Sulfur.} Next we considered shifting the N
impurity states closer to the valence band by proper alloying of MgO
with a third component. In this respect, MgO$_{1-y}$S$_y$ seemed
promising, because MgS also crystallizes in the rock-salt structure,
with a larger lattice parameter (5.2 \AA) and a smaller, indirect band
gap (experimental: 4.5~eV ; we find 2.25~eV within the LDA).
Calculations in MgO$_{1-y}$S$_y$ (which were carried out by changing
the lattice constant according to Vegard's law) showed that the O
valence states end up higher than the S valence states (see
Fig.~\ref{fig:dos}(b)). This result is counter-intuitive if one takes
into account only the electronegativity of the elements. We interpret
it by observing that the stronger localization of the O and N $2p$
states is responsible for a stronger on-site Coulomb repulsion, so
that filled $2p$ levels end up higher in energy than the more extended
S $3p$ levels. As a consequence, the valence band edge in
MgO$_{1-y-x}$S$_y$N$_{x}$ is again Oxygen-dominated, but the 
position of the N impurity band is closer to the Oxygen valence band
edge, and there is a stronger hybridization of the N with the O states
compared to MgO$_{1-x}$N$_{x}$, as can be seen in
Fig.~\ref{fig:dos}(a,b). Because of this, the pair exchange constants
are more extended in MgO$_{1-y-x}$S$_y$N$_{x}$. Calculations of
MgO$_{0.45}$S$_{0.5}$N$_{0.05}$ showed an enhancement of $J(R)$ at
large distances by up to 10-20\% compared to
MgO$_{0.95}$N$_{0.05}$. Thus the Curie temperature increases by a
factor of the same order, remaining small.

--\emph{Doping with holes.} In principle, the double-exchange
mechanism for ferromagnetism is most effective when the impurity band
of one spin is exactly half-filled, while for the opposite spin is
completely full (or empty). Then the energy gain in the ferromagnetic
state by band-broadening is maximized. In MgO$_{1-x}$N$_{x}$ this is
not the case, as the N spin-down band is filled by 2/3. Therefore,
doping with holes could help, as \EF\ will be shifted downward to the
middle of the impurity band. Note that we are not trying to achieve
the analogue of Zener's hole-mediated $p$-$d$ exchange, but rather an
improvement of the double-exchange mechanism. (Similar engineering has
been proposed in transition-metal-doped DMS.\cite{Fukushima06}) We
investigate this effect in a Mg-poor compound i.e.,
Mg$_{1-y}$O$_{1-x}$N$_{x}$. Each missing Mg atom adds two holes; thus
the ``ideal'' Mg-vacancy concentration, resulting in exactly
half-filling of the N spin-down band, is $y=x/4$. In our calculations
we used $x=10\%$ and $y=2.5\%$. As it turns out, the exchange
interactions become more extended in space, however, the reason is
that \EF\ now slightly enters the O valence band (since the valence
band top is anyhow merged with the lower part of the impurity band),
and a Fermi surface is now available for the impurity interaction. As
a result, a  RKKY mechanism emerges,
with an outcome of also antiferromagnetic interactions at large
distances. The latter turn out to be comparatively strong, once more
because of an additional increase of superexchange, as \EF\ shifts
closer to the spin-up band. Thus there is no increase of $T_{\rm C}$. Tests
with lower hole-doping concentration, before the RKKY exchange sets
in, result in only a marginal increase of the long-range interactions,
insufficient for a significant increase of $T_{\rm C}$.

\section{Discussion on approximations \label{sec:approximations}}

\subsection{Approximations in the calculation of the electronic
  structure and Curie temperature}

We now consider which approximations used in our calculations could most
seriously affect the electronic structure, exchange constants and
Curie temperature. Perhaps the most critical is the local density
approximation. Corrections that could qualitatively change the
electronic structure include self-interaction correction, inclusion of
strong on-site Coulomb repulsion, or $GW$-type of self-energy.

In Ref.~\onlinecite{Droghetti08}, Droghetti and Sanvito applied the
self-interaction correction (SIC) to MgO doped with N (among other
compounds). They find that an application of SIC to both N and O atoms
increases the gap, as is expected by driving the occupied levels lower
by the SIC. However, their main finding connected to our present
discussion is the occurrence of a strong splitting of the order of
3~eV between the occupied and unoccupied $p$ states of N. The
spin-down impurity band is, after SIC, manifestly insulating. A
similar effect was found within the LDA+$U$ method, applied by Pardo
and Pickett\cite{Pardo08} to a number of oxides (including MgO) with N
substitution of Oxygen. In the latter work, a Coulomb repulsion of
$U=5.5$~eV was used on the O and N atoms. From the point of view of
ferromagnetism, an insulating impurity band can have severe
consequences, in particular disfavoring the double-exchange mechanism
(impurity-band broadening does not lead to energy gain any more);
ferromagnetic superexchange could be present, but it is much
weaker. In these cases, the system could even show a spin-glass ground
state instead of a ferromagnetic one.

These works demonstrate possible consequences of strong Coulomb
interactions, showing that local density-functional theory could be
insufficient. We believe that, at low concentrations, these results
undoubtedly show the correct physics, but at higher concentrations
(a few percent) they possibly overestimate the Coulomb interaction, and
are therefore still inconclusive as to the exact nature of the ground
state. Our arguments are as follows.

First, considering the low-concentration limit, the N impurity states
are not localized at the N atomic cell, but rather extended also over
the 12 nearest Oxygen neighbors, as our calculation shows. This speaks
for a relatively mild self-interaction error, compared to $d$- or
$f$-systems, where the SIC is usually applied to. The LDA error must
become less and less serious as the concentration increases and an
itinerant impurity band is formed. Furthermore, application of Coulomb
correction terms on the Oxygen valence band is probably much less
necessary, as the associated Bloch states are itinerant with a
significant band width (in the limiting case of completely
non-localized states of a homogeneous electron gas, the
self-interaction is fully corrected by the LDA exchange-correlation
energy). We conclude that Coulomb-interaction corrections should be of
different scale at the N impurity-band states compared to the O
valence-band states, but even for N they should not be too severe.

Second, as the concentration $x$ increases, the impurity band width
$w\sim\sqrt{x}$ increases rather fast, exceeding 1~eV already at
$x=5\%$. This band width can well be of the order of magnitude of the
local Coulomb interaction $U$ (in the LDA+$U$ calculations of
Ref.~\onlinecite{Pardo08}, a value of $U=5.5$~eV was used, which,
however, acts on the atomic site; if the full extent of the impurity
wavefunction is accounted for, a significantly lower value of $U$
would be needed for the same effect). In this case a Mott transition
is possible, from an insulating state for low concentrations (where
$w<U$) to a metallic state at high concentrations (where $w>U$). In
such a scenario, ferromagnetism would be assisted by the increase of
concentration in two ways: occurrence of a metallic state similar to
the LDA result, and magnetic percolation.

Concerning $GW$-type of corrections, these are known to correct the
LDA underestimation of band gap in band insulators. It should be then
expected that the larger gap would lead to a stronger decay of the
exchange constants with distance, reducing the calculated
$T_{\rm C}$. However, this reduction should not be too serious, for the same
reason that $T_{\rm C}$ cannot be much increased by engineering a smaller gap
(see Sec.~\ref{sec:change}).

Another approximation that was made here was the assumption of a
classical, rather than quantum, Heisenberg model to describe the
magnetic excitations. Considering that we are faced with an
$S=\frac{1}{2}$ system, it can be argued that a classical
approximation is unrealistic. Within the RPA, a change from a
classical to a quantum Heisenberg model leads to an increase of $T_{\rm C}$
by a factor of $S(S+1)/S^2=3$, in the $S=\frac{1}{2}$ case. Accepting
this, room-temperature ferromagnetism could be achieved at $x=20\%$,
as can be seen by scaling up the results of
Fig.~\ref{fig:Tc}. However, the particular way of calculation of the
exchange coupling constants\cite{Liechtenstein87} $J_{nn'}$ tacitly
assumes a classical model. The constants are calculated within
constrained density-functional theory, in principle by ``freezing''
the system in a static non-collinear configuration and calculating the
total energy; this should be viewed as a parametrization of the
low-energy excitations of the spin density, rather than a derivation
of a quantum spin Hamiltonian. Comparison to experiments in previous
works also advocates for this point of view. E.g., Sasioglu et
al.\cite{Sasioglu05} have calculated the Curie temperature of Heusler
alloys by a similar recipe. While the assumption of a classical Heisenberg
model lead them to reasonable agreement of $T_{\rm C}$ with experiment, the
assumption of a quantum model resulted in a clear overestimation of
$T_{\rm C}$. For lack of a better theory that predicts $J_{nn'}$ for a
quantum Heisenberg model, we are obliged to work within a classical
model.

\subsection{Approximations imposed by our structural model}

So far we assumed a uniform, on the average, distribution of the N
atoms in the MgO matrix.  We therefore start this Subsection by
commenting on the solubility of N in MgO. The solubility limit of
substitutional N is expected to be small, and clustering of the N
impurities is expected to be favored. We verified this by calculating,
within the KKR-CPA, the mixing energy as a function of the
concentration, $E_{\rm mix}(x) = E[\text{MgO}_{1-x}\text{N}_x] -
(1-x)E[\text{MgO}] - x E[\text{MgN}]$ with MgN in the rock-salt
structure. $E_{\rm mix}(x)$ was positive for all calculated
concentrations ($1\%\le x\le 90\%$), showing that phase separation
into MgO and MgN is energetically favored. This can be understood also
from the local density of states, shown in Fig.~\ref{fig:dos}(a),
where it is seen that the impurity band is bisected by the Fermi level
\EF. In case of impurity clustering the bandwidth will increase and
energy will be gained because of the lowering of the occupied
levels. Interestingly, the origin of ferromagnetic double exchange
(bisection of the impurity band by \EF) provides also a mechanism for
(usually unwanted) phase separation. Of course, phase separation other
than the MgO--MgN type could be also possible; what this calculation
shows is the thermodynamic instability of MgO$_{1-x}$N$_x$ with
respect to at least one type of phase separation, even if entropic
effects could milden the separation. The preference toward clustering
was also seen in calculations of two substitutional N impurities in a
MgO supercell matrix: the most stable state was found when the two N
atoms were first neighbors in the $\langle 110\rangle$ direction.  We
conclude that MgO$_{1-x}$N$_x$ can only be grown under
out-of-equilibrium conditions, as is the case with many
transition-metal-doped DMS.

Two neighboring substitutional N atoms still provide a metallic
state. However, three neighboring N atoms forming an equilateral
triangle in a \{111\} plane cause a transition to a semiconducting
state (with a split minority-spin band), if structural optimization of
the three atoms and their surroundings is taken into account.

Coming now to interstitial N impurities, our calculations (performed
with VASP) showed that strong atom displacements are involved, thus we
cannot use CPA to describe the random alloy with interstitials. As the
phase space involved is extremely rich, and probably depends strongly
on growth and annealing conditions, we defer a deeper analysis of the
particular question to a future work, commenting on a few interesting
findings.

First, in the case of a single N interstitial, the symmetric,
tetrahedral position constitutes only a local minimum of the total
energy. Further significant reduction of the total energy is found in
a ``dumbbell'' configuration reminiscent of a NO molecule, where the N
impurity binds itself to an O atom, with both atoms in the proximity
of the ideal O lattice position. This entity is found to be
half-metallic and magnetic, with a moment of 1~\muB.

Second, when two Nitrogen atoms are placed around an Oxygen atom as
interstitials, structural relaxation leads to a ``zig-zag'' O-N-O-N
configuration. Here, the two N atoms are almost on the lattice sites,
being first neighbors along $\langle 110\rangle$, while the O atoms
are shifted out of their ideal positions, hovering above the $(110)$
atomic layer. The electronic structure changes to non magnetic and
insulating.

Third, the following configuration is of particular interest: two
neighboring N atoms together with an O interstitial between them, plus
a nearby O vacancy. This is so to say a configuration where an O atom
has been bound during growth by a N pair, missing the lattice position
nearby. This configuration shows a local energy minimum and is
non-magnetic. However, if the O atom returns to the vacant lattice
position, the state becomes magnetic with a total energy gain of the
order of 4~eV, while the lattice parameter is reduced by 1\% (in the
particular calculation we took a Mg$_{32}$O$_{30}$N$_2$ supercell,
corresponding to 6.25\% N). This is a possible explanation of the
appearance of magnetism after annealing in experiment.\cite{Parkin}

\section{Conclusions and outlook \label{sec:conclusions}}

The chemical compound MgO$_{1-x}$N$_x$ in which Oxygen is substituted
by Nitrogen atoms is found to be ferromagnetic for all concentrations
which seem realistically achievable ($x\leq 15\%$) and above the
concentration of the percolation threshold (about 1.5\%). The N atom
forms a partially occupied gap state at the vicinity of the valence
band edge. This leads to concentration-dependent exchange parameters,
exponentially decreasing with distance, which are ferromagnetic due to
double exchange as dominating mechanism. The reduction of the average N-N
distance and the reduction of the exchange interaction strength with
increasing concentration compensate partially, so that the Curie
temperature increases linearly with concentration. The structural
relaxation around the N impurity are found to be small.  In the case
of N clusters and interstitial N we found magnetic as well as
non-magnetic complexes, but never with higher moments than 1~\muB/atom.

We think that the Curie temperature of such $sp$-compounds can be
increased if the exchange interactions become more long-ranged, e.g.
by moving the gap states closer to the band edge, and if the N-N
exchange interactions are enhanced by suppressing the competing
antiferromagnetic superexchange via an increase of the local exchange
splitting. Our attempts in this direction, using compounds such as
MgO$_{1-x-y}$S$_y$N$_x$ or Mg$_{1-y}$O$_{1-x}$N$_x$, have not been met
with success. Nevertheless, it is worthwhile to investigate more
host/impurity combinations theoretically, in order to guide
experimental efforts.

From the point of view of applications, a low $T_{\rm C}$ is a
limiting factor. However, from the point of view of physics, $sp$
magnetism in oxides presents intriguing open problems, such as the
nature of the ground state, its concentration dependence, and the role
of dynamic electron correlations. Moreover, although homogeneous
Nitrogen distributions at high concentrations are difficult to
achieve, it is worthwhile to examine the physics and technological
relevance of inhomogeneous samples, emerging, e.g., by spinodal
decomposition\cite{Fukushima07} or delta-doping,\cite{Picozzi06} as
has been proposed for transition-metal doped DMS. Such compounds could
show much more robust magnetism locally, leading to new
functionalities.

\appendix

\section{Methods of calculation}

For the calculations of the alloy ground-state electronic structure
(except structural relaxations) we used the full-potential
Korringa-Kohn-Rostoker Green-function method\cite{SPRTBKKR} (KKR) with
exact calculation of the atomic cell shapes,\cite{Stefanou91} using an
angular momentum cutoff of $l_{\rm max}=3$ and a maximal $k$-point
mesh of 64000 points in the full Brillouin zone; the complex-energy
contour integration included the Oxygen and Nitrogen $2s$
states. Disorder was treated within the Coherent Potential
Approximation (KKR-CPA). Relativistic effects were neglected. In the
KKR-CPA calculations we used the MgO experimental lattice parameter
for all concentrations, as the small change should have no significant
effect in the calculated trends.

For the calculation of the hole localization, the impurity-in-host
version of the KKR method was employed non-self-consistently, with
potentials taken from self-consistent CPA. In the KKR-CPA
calculations, the LDA for the exchange-correlation energy was used
with the parametrization of Vosko et al.\cite{Vosko80} For the structural
relaxations we used the Vienna {\em ab-initio} simulation package\cite{VASP}
(VASP) within the GGA.\cite{GGA}

The exchange constants $J_{nn'}$ were calculated within the
approximation of infinitesimal rotations,\cite{Liechtenstein87} again
within the KKR-CPA. As has been found by Sato et al.,~\cite{Sato06}
the method of infinitesimal rotations within the CPA can overestimate
the values of $J_{nn'}$ between the nearest neighbors at low
concentrations (when they anyhow are irrelevant for $T_{\rm C}$), but
is rather good for larger distances.

The Curie temperatures were calculated within the random-phase
approximation (RPA) for disordered systems, as proposed by Hilbert and
Nolting,\cite{Hilbert04} with the distance-dependent interactions
$J_{nn'}$ taken as described above. At each concentration, an
environmental average was evaluated by taking 100 random
configurations of approx.~8500 nitrogen atoms each, statistically
distributed in a simulation supercell (sized between $21^3$ unit cells
for 90\% concentration and $94^3$ unit cells for 1\%
concentration). To each configuration the RPA yielded a $T_{\rm C}$,
and an average of all values of $T_{\rm C}$ was determined at the
end. In special cases ($x=5\%$, 10\%, 15\% and 10\%) the results were
cross-checked with Monte Carlo calculations, which yielded a $T_{\rm
  C}$ of approximately 50\% higher, except for 5\% where the results
of the two methods agreed. Because the Monte Carlo method is
numerically more expensive, in these tests we treated systems of
1000-2500 magnetic ions, using up to 20 configurations for averaging;
within Monte-Carlo, $T_{\rm C}$ was calculated using the 4th
order cumulant method.\cite{Binder}

\acknowledgments 

We would like to thank Stuart Parkin for discussions on his
experimental results prior to publication. We are grateful to
Peter H. Dederichs for his constant support and enlightening
discussions. This work was funded in part by the Young Investigators
Group Programme of the Helmholtz Association (``Computational
Nanoferronics Laboratory,'' Contract VH-NG-409).

%%%%%%%%%%%%%%%%%%%%%%%%%%%%%%%%%%%%%%%%%%%%%%%%%%%%%%%%%%%%%%%%%%

\end{document}